\documentclass[%
 aip,
 amsmath,amssymb,
 reprint,%
]{revtex4-1}

\usepackage{graphicx}
\usepackage{dcolumn}
\usepackage{bm}
\usepackage{soul}
\usepackage[usenames,dvipsnames]{color}

\usepackage[utf8]{inputenc}
\usepackage[T1]{fontenc}
\usepackage{mathptmx}
\usepackage{etoolbox}

\newcommand{\nb}[1] {{\color{black}#1}}

\makeatletter
\def\@email#1#2{%
 \endgroup
 \patchcmd{\titleblock@produce}
  {\frontmatter@RRAPformat}
  {\frontmatter@RRAPformat{\produce@RRAP{*#1\href{mailto:#2}{#2}}}\frontmatter@RRAPformat}
  {}{}
}%
\makeatother
\begin{document}

\preprint{AIP/123-QED}

\title{Computational inverse design of acoustoplasmonic metasurfaces}
\author{Julia E. Holland}

\affiliation{ 
Department of Mechanical $\&$ Aerospace Engineering, UC San Diego, La Jolla, California 92093-0021, United States
}%

\author{Nicholas Boechler}%

\author{Lisa V. Poulikakos}
\email{lpoulikakos@ucsd.edu}
\affiliation{
Department of Mechanical $\&$ Aerospace Engineering, UC San Diego, La Jolla, California 92093-0021, United States
}

\affiliation{
Program in Materials Science $\&$ Engineering, UC San Diego, La Jolla, California 92093-0021, United States
}

\date{\today}

\begin{abstract}
Optical and acoustic metasurfaces are two-dimensional arrays of subwavelength elements that locally modulate or phase shift incident waves. Acoustoplasmonic metasurfaces combine the physics of light and sound, producing acoustic wavefronts in response to optical stimuli. Herein, we present a computational inverse acoustoplasmonic metasurface design algorithm for desired optically-generated acoustic wave fields. We consider gold nanoparticles producing spherical acoustic waves in water, and the resulting acoustic wave propagation along the plane containing the nanoparticle array. We demonstrate how our algorithm can be used to design metasurfaces that can be used to achieve complex acoustic wave fields. This includes the design of a single metasurface that produces acoustic wave fields mimicking two different Morse code patterns upon stimulation with two orthogonal polarization states of light. This work provides a new tool for the design of complex optically generated acoustic wavefronts, enabling functionality beyond what would be achievable with off-optical-resonance optoacoustic excitation.  
\end{abstract}

\maketitle

%


Metasurfaces comprise two-dimensional (2D) arrangements of sub-wavelength structures which can manipulate the interaction of waves (e.g. optical or acoustic waves) with matter. \cite{kuznetsov2024roadmap,assouar2018acoustic,ji2022recent} Various inverse design methods, including combinations of analytical and computational tools, have been used to design metasurfaces that provide translation between an input and desired output wave field, enabling versatile and tunable wavefront shaping. Such metasurface design algorithms are especially prevalent in holography -- both optical\cite{zou2023metasurface,elsawy2020numerical} and acoustic\cite{melde2016holograms,miao2023deep} alike.\cite{liu1997propagation} Another common application for such metasurface inverse design is for focusing waves through scattering media, again in both optics\cite{fayyaz2019comparative,haim2024image} and acoustics.\cite{haddadin1998ultrasonic}

\begin{figure*}
\includegraphics[width=6.69in]{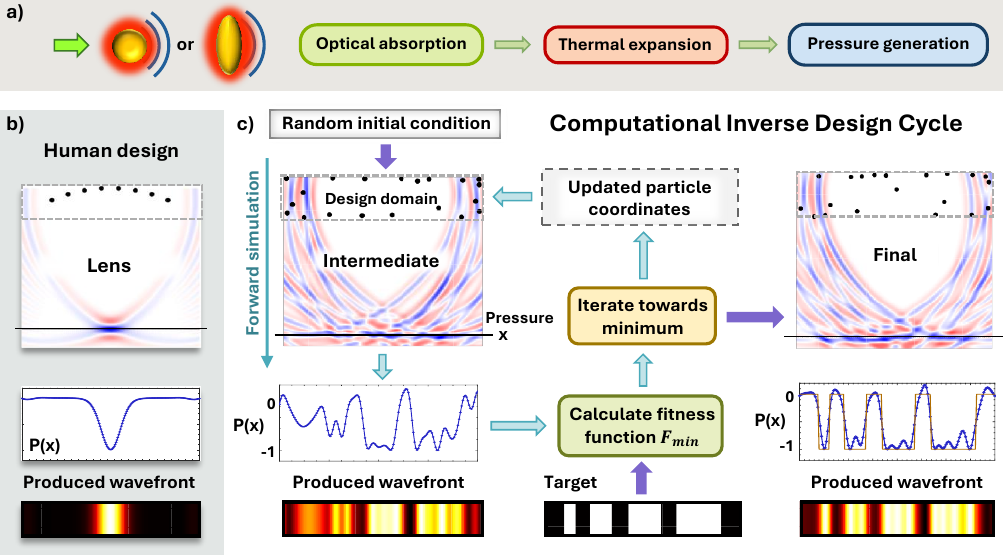}
\caption{\label{fig:schematic} \textbf{Computational inverse design of plasmonic particle coordinates to achieve desired acoustic wave fields.} (a) Schematic showing conversion of
absorbed light to mechanical vibration via photothermal heating of a plasmonic particle. (b) A circular source array of plasmonic particles producing a focused, Gaussian beam profile at the focal plane. (c) Computational inverse design cycle used to identify 2D plasmonic particle arrays that produce a desired acoustic wave field. 
}
\end{figure*}

Acoustoplasmonic metasurfaces combine the physics of light and sound.\cite{berte2018acoustic,imade2021gigahertz,poblet2021acoustic,boggiano2024focusing} Therein, sub-optical-wavelength plasmonic nanoparticles strongly absorb incident light via excitation at or near their localized surface plasmon resonance (LSPR). When excited by pulsed light, the particle will mechanically vibrate as a result of rapid heating and subsequent thermoelastic strain, leading to pressure generation in the surrounding medium (Fig. \ref{fig:schematic}a). This mechanism parallels that of laser ultrasonics, in which bulk material or thin films (common examples) are irradiated with laser light, generally off optical resonance.\cite{drain2019laser} In the case of thin films, surface plasmon polaritons can be activated over a larger surface area,\cite{temnov2016towards} in contrast to the LSPR optical confinement in nanoparticles. In laser ultrasonics, wavefront shaping has been achieved by spatially manipulating incident irradiation, \cite{cielo1985laser,bruno2016laser} or, more closely, through the optical excitation of metallic patterns. \cite{matsuda2020optical} 
In acoustoplasmonics, recent experimental demonstrations of surface acoustic wave focusing were achieved via the near-LSPR optical excitation of nanoparticles which were intelligently positioned to generate tailored acoustic wavefronts.\cite{boggiano2024focusing} One key advantage of the use of LSPRs over metallic patterns excited off the plasmon resonance, is that the LSPRs introduce additional tunability via effects such as optical polarization sensitivity, arising from optical anisotropy in the nanoparticle geometry or material composition. \cite{holland2024acoustoplasmonic} Moreover, acoustoplasmonics involves the interplay of optical and acoustic resonances, tunable via the electromagnetic frequency of the incident light and the excitation laser pulse duration, respectively. \cite{holland2024acoustoplasmonic} However, algorithmic optimization is required to fully leverage the versatile degrees of freedom that acoustoplasmonic metasurfaces enable for tunable acoustic wavefront shaping.

Herein, we present a computational inverse acoustoplasmonic metasurface design algorithm for desired optically generated acoustic wave fields. We consider gold nanoparticles producing spherical acoustic waves in water, and the resulting acoustic wave propagation along the plane containing the nanoparticle array. Using spherical nanoparticles, we first test our algorithm against the simple case of a circular equispaced source array that produces a Gaussian pressure spatial profile along a line formed by the intersection of the plane containing the nanoparticles and the focal plane (Fig. \ref{fig:schematic}b). Second, using ellipsoidal nanoparticles, we use our algorithm to achieve two complex wave fields mimicking two Morse code patterns. In this second case, we show how the polarization-dependent, LSPR-enabled optical absorption of nanoellipsoids can be used to produce a polarization-switchable one-dimensional (1D) holographic metasurface design. We suggest that such polarization tunability provides an example of advantages that may be obtained via acoustoplasmonic wavefront generation, wherein such functionality would not be achievable for metasurfaces involving solely scalar acoustic wavefields or off-optical-resonance optoacoustically excited particle arrays. 


Figure \ref{fig:schematic}c provides a schematic of the inverse design algorithm. We use the \textit{fmincon} function in MATLAB, which finds the minimum of a constrained nonlinear multivariable function. 
Initial conditions comprise randomized particle coordinates within a user-defined design domain (indicated by the gray dashed-line box). The fitness function is computed by comparing a calculated (``intermediate'') and target 1D pressure profile along a line defined in the plane containing the nanoparticles. The calculation of the intermediate pressure profile from given particle positions constitutes the ``forward simulation'' part of the cycle. The \textit{fmincon} function then adjusts the particle coordinates until a local minimum in the fitness function is found (default step tolerance of $1\times 10^{-10}$), or when the optimizer exceeds the maximum number of allowed iterations or function evaluations (left at their default values of $1\times 10^3$ and $3\times 10^3$, respectively). We include an additional constraint which ensures a minimum particle center-to-center distance of 60 nm. The resulting particle coordinates produce the ``final'' pressure profile. Taking into account the non-convexity  of the design space, this process is repeated for multiple random initial conditions, using the particle configuration resulting in the smallest fitness function indicating the final solution.

In the forward simulation, the nanoparticles are excited by a Gaussian laser pulse (20 ns pulse duration $\tau_P$). When considering nanospheres, since we are interested in normalized pressure magnitude profiles, we choose an arbitrary nanosphere absorption efficiency (refer to Appendix \ref{sec:analytical}) such that wavelength is not considered. For the case of nanoellipsoids, the incident optical wavelength is chosen to be 916 nm. This optical pulse excites broadband, sub-mechanical resonance acoustic waves, which we approximate as point-sources. The frequencies excited are sufficiently low to neglect acoustic attenuation in the fluid.\cite{kinsler2000fundamentals} The temporal pressure response at the particle surface is analytically calculated for the case of nanospheres (refer to Appendix \ref{sec:analytical} for derivations) and numerically calculated via finite element method (COMSOL Multiphysics) for nanoellipsoids (refer to Appendix \ref{sec:numerical} details). \nb{The pressure generated at each nanoparticle is then propagated analytically and superimposed to obtain the resulting pressure field.} \cite{holland2024acoustoplasmonic}

\begin{figure*}
\includegraphics[width=5.9in]{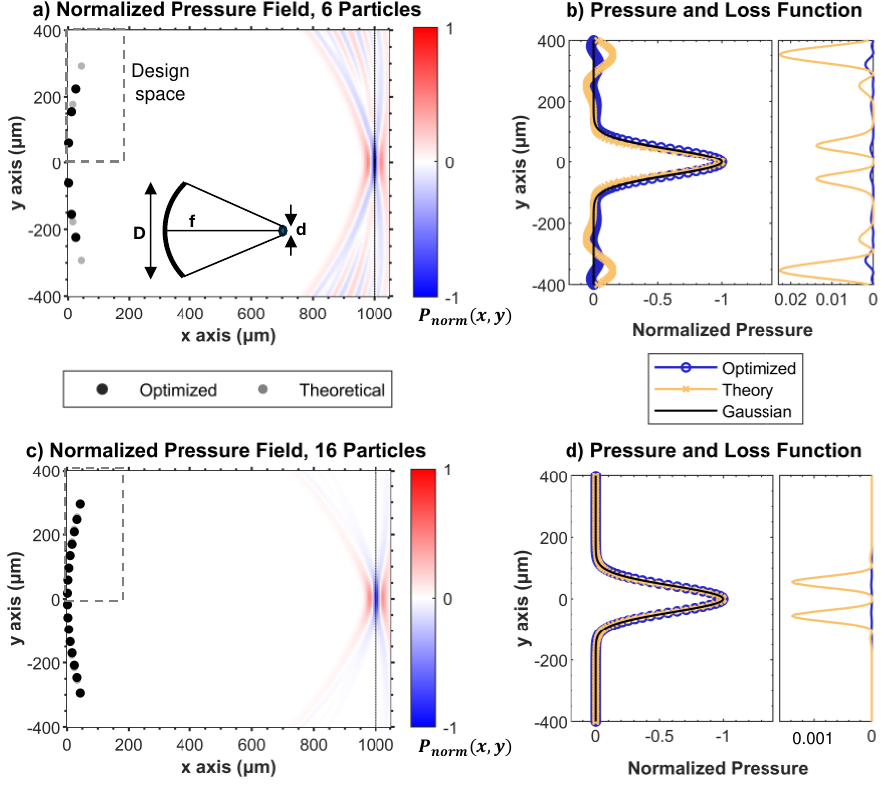}
\caption{\label{fig:lens} \textbf{Comparison of equispaced circular particle array and optimized particle arrangement for focusing into a Gaussian pressure profile.} Top row: 6 particles. Bottom row: 16 particles. (a,c) Normalized pressure field generated from optically excited nanospheres having optimized coordinates. Particle locations are indicated by the black circles (diameters not to scale). For reference, the theoretically calculated particle coordinates are indicated by the gray circles. The design space is noted by the gray dashed line box, and the focal length is noted by the dotted black line at $x = 1000$ $\mu$m. (a, inset) Schematic of ideal lens. (b,d) Left: Normalized pressure magnitude at $x = f$ generated from the optimized (blue, circle) and equispaced circular array (gold, x) particle coordinates with the Gaussian curve (black line) shown for reference. Right: Loss function (Eq. \ref{eq:loss}) of the optimized (blue) and equispaced circular array (gold) particle coordinates.}
\end{figure*}

As noted, we first compare the performance of our inverse design algorithm to that of the pressure profile produced by a circular source array, mimicking the behavior of a lens (Fig. \ref{fig:lens}). \nb{We note that, in a sense, this circular array can be considered a sort of transducing lens, in that a plane optical wave is transformed into a focused acoustic wave field.} The circular array is chosen to have a 1000 $\mu m$ focal length $f$ and 100 $\mu m$ full width half maximum (FWHM) focal spot $d$, for the pressure along the line defined by the focal plane.  Considering Abbe's diffraction limit,\cite{hecht2012optics} we can obtain the diameter on the lens $D$ (the array chord), as a function of the focal length and FWHM via
\begin{equation}
    D=2f\tan\left[ \sin^{-1}\left(\frac{\lambda}{2d} \right)\right],
\end{equation}
where we choose $\lambda$ via Fast Fourier Transform as the peak acoustic wavelength of the excited nanoparticle pressure response (see Fig. \ref{fig:lens}a inset). 
The nanoparticles are spaced evenly along the circular array. 
The wave field is evaluated at $t=v/f$, where $v$ is the speed of sound in water \nb{and time $t=0$ is set as the time of the maximum pressure magnitude excited at the particle surface}. 
For this case of a circular array, we take advantage of symmetry over $y = 0$, choosing an even number of particles and optimizing over a design domain space bounded between $0<x<200$ $\mu$m and $0<y<400$ $\mu$m (dashed gray box in Fig. \ref{fig:lens}a,d).

For the computational inverse lens design, we define our fitness function as
\begin{eqnarray}
F_{min}&=&\sqrt{\sum{\left[\left(\frac{P(f,y)}{\text{max}\left(|P(f,y)| \right)}-G(y) \right)^2\right]}}\nonumber \\
 & &+ \frac{2RN_p}{f}
  \frac{\text{max}\left(|P_{t=0}|\right)}{\text{max}\left(|P(f,y)| \right)}~,
\end{eqnarray}
where $P(x,y)$ is the pressure response, $G(y)$ is the Gaussian pressure shape, $R=20$ nm is the particle radius, $N_p$ is the number of particles, and $P_{t=0}$ is the pressure excited at the particle surface. The first term in this fitness function rewards the pressure profile that matches the Gaussian shape, and the second term rewards a maximum pressure magnitude at the focal point. This second term is scaled by a factor of two to prioritize this objective. Practically, this ensures that the optimizer does not ``throw away'' particles.

Figure \ref{fig:lens} compares the pressure field generated by the circular, equispaced particle array with that of an optimized particle arrangement. We show this comparison for an array comprising 6 particles (Fig. \ref{fig:lens}a,b) and 16 particles (Fig. \ref{fig:lens}c,d). Figure \ref{fig:lens}a and c show the normalized pressure fields at $t=v/f$ generated from optimized particle coordinates (black circles), which are compared to the circular array coordinates (light gray circles). The normalized pressure profiles across the y axis at $x = f$ are compared for the equispaced circular and optimized arrangements in Fig. \ref{fig:lens}(b,d, left). The Gaussian curve (black solid line) is shown for reference. 
The optimized particle arrangement can be seen to not only achieve focusing, but surpasses the equispaced circular array in regards to approaching the target Gaussian spatial profile for both the 6 and 16 particle arrays. We further demonstrate this improvement via a loss function defined as
\begin{equation}
    L = \left(\frac{P(f,y)}{\text{max} \left(|P(f,y)| \right)} - G(y)\right)^2,
    \label{eq:loss}
\end{equation}
which we plot in Fig. \ref{fig:lens}(b,d, right). 

\begin{figure}
\includegraphics[width=3.3in]{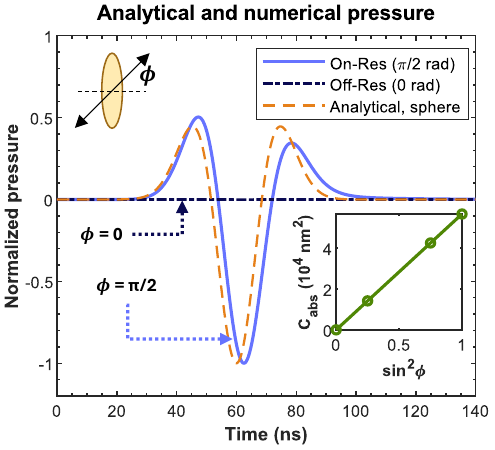}
\caption{\label{fig:P0} \textbf{Particle anisotropy enables optical polarization-switchable pressure generation.} Normalized numerical pressure generation at the nanoellipsoid surface (probed at the minor axis) for light linearly polarized along the major (solid light purple line) and minor (dark purple dashed line) axes, \textit{i.e.}, on and off-resonant optical excitation, respectively. These are also denoted by $\phi=\pi/2$ and $\phi=0$, respectively, where $\phi$ is the polarization angle measured from the horizontal for a vertically-oriented nanoellipse (bottom left schematic). The normalized analytical pressure generation calculated for a sphere is plotted for reference (orange dashed line). Bottom right inset: absorption cross-section plotted over $\sin^2\phi$.}
\end{figure}

We next consider the ability to create optical-polarization-tunable pressure profiles. To do this, we use gold nanoellipsoids with a minor axis radius of 20 nm and a major axis radius of 80 nm. The anisotropic particle geometry results in polarization-tunable optical absorption efficiency and subsequent pressure generation.\cite{lanzillotti2018polarization,holland2024acoustoplasmonic} Figure \ref{fig:P0} demonstrates this phenomenon via the pressure generated with an on- (light purple, solid line) and an off-resonant optical excitation (dark purple, dashed line), showing near zero pressure generated for the off-resonant case. At the incident wavelength, on resonant excitation occurs when light is linearly polarized along the major axis, and off-resonant excitation when polarized along the minor axis. This corresponds to $\phi = \pi/2$ and $\phi = 0$, respectively, where $\phi$ is the polarization angle measured from the minor axis. The inset shows the absorption cross-section $C_{abs}$ (geometric cross-section scaled by the absorption efficiency) plotted over $\sin^2\phi$. 
The analytically derived pressure, normalized by its peak amplitude, (yellow dashed line) for a nanosphere ($R = 20$ nm) is shown for reference, the temporal profile of which agrees well with the simulated on-resonance pressure temporal response for a nanoellipsoid, normalized by its peak amplitude, even with the elongation of the major axis.

\begin{figure*}
\includegraphics[width=6.69in]{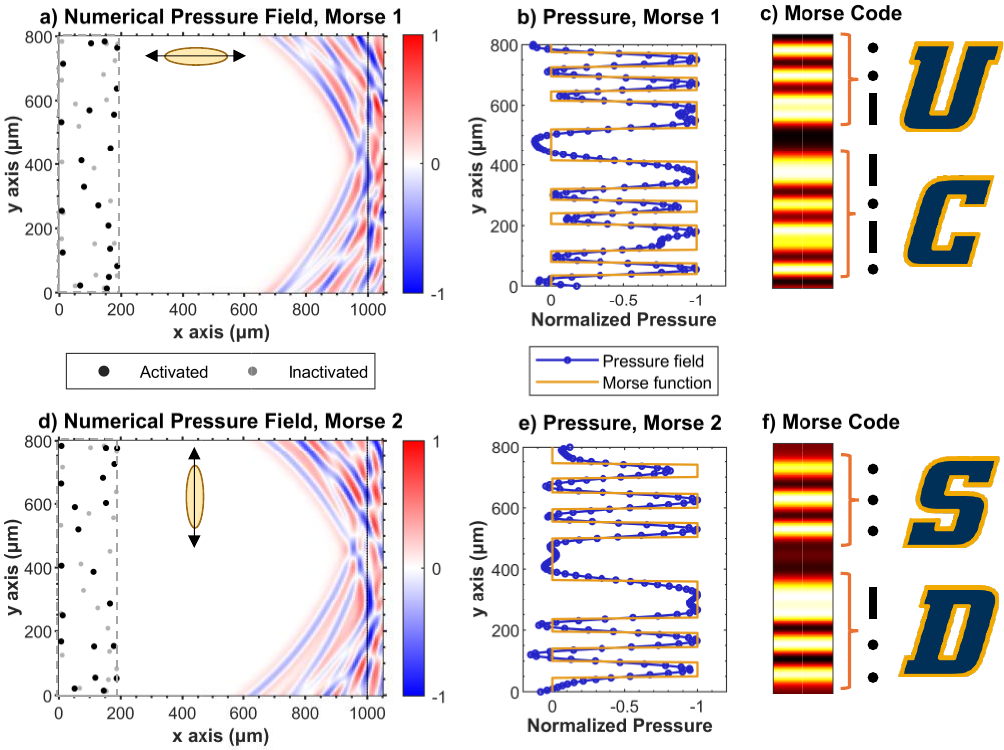}
\caption{\label{fig:morse} \textbf{Nanoellipsoid particle arrangements for polarization-switchable acoustic Morse code profiles.}  Pressure generated at $t=f/v$ via the excitation of horizontally- and vertically-oriented nanoellipsoids with, respectively, horizontally- and vertically-polarized light (top and bottom rows, respectively). a,d) Normalized pressure fields. Insets: nanoellipsoid  orientation and light polarization direction (arrow). b,e) normalized profile at $x = f$ generated from the optimized (blue, circle) particle coordinates; the Morse code pressure target (gold solid line) is shown for reference. c,f) Unsaturated heat map of the achieved normalized pressure at $x = f$ and the Morse code letter equivalent.}
\end{figure*}

In our second example, we use the optical-polarization-dependent excitation of the gold nanoellipsoids to design a single 40 particle array that creates one pressure profile with vertical polarization and a second pressure profile for horizontal polarization. The near-zero, off-optical-resonance pressure response allows us to split the problem into two parts. We design one pattern for 20 vertically oriented nanoellipsoids and one pattern for 20 horizontally oriented ellipsoids and then superimpose into a single array. Figure \ref{fig:morse} demonstrates our method's ability for polarization-switchable wave tailoring in the form of a single acoustoplasmonic metasurface capable of producing two distinct pressure profiles in the form of Morse code patterns. 
Figure \ref{fig:morse}a and d show the wave fields generated from nanoellipsoids oriented horizontally and vertically, respectively. Both particle sets are shown in each plot, with the activated particles shown as larger black dots and the inactivated as smaller gray dots. Despite the change in particle geometry and orientation, the pressure propagation can still be approximated as that of a spherical wave in the acoustic far field.\cite{wang2022understanding}

The fitness function for each pattern in this example is defined as
\begin{eqnarray}
F_{min}&=&\sqrt{\sum{\left[\left(\frac{P(f,y)}{\text{max}\left(|P(f,y)| \right)}-M(y) \right)^2\right]}}\nonumber \\
 & &+ \frac{1}{N_{peaks}}\frac{RN_p}{f}
  \frac{\text{max}\left(|P_{t=0}|\right)}{\text{max}\left(|P(f,y)| \right)}~,
\end{eqnarray}
where $M(y)$ is the user-defined Morse code profile and $N_{peaks}$ is the number of peaks in the pressure profile --- \textit{i.e.}, the number of Morse symbols, whether a dot or a dash. In $M(y)$ the ``on'' signal corresponds to a normalized pressure of $-1$ and ``off'' to zero. The spatial length in $y$ along the pressure profile measurement line at $x=f$ from the prior example differentiates dots and dashes. In designing these Morse code functions, we note that the smallest design feature, \textit{i.e.}, a ``dot'', should not be smaller than $d = v\tau_P$.

Figure \ref{fig:morse}b and e compare the resulting normalized pressure profiles (solid, dotted purple line) and the respective Morse code profiles (solid gold line) for each polarization, which give good agreement. These results are represented as unsaturated heat maps in Figure \ref{fig:morse}c and f, in which the ``dots'' and ``dashes'' are clearly visible. The corresponding letters are also shown for reference.

In summary, we have demonstrated a computational inverse design method to generate plasmonic particle arrangements in a 2D domain to produce a desired 1D acoustic wave profile upon optical excitation. \nb{We suggest that more complex wavefronts and subtle tunability could be enabled by considering non-binary polarizations and particle angles, as well as particle shape optimization.}  This includes potential use of chiral acoustoplasmonic nanoparticles and circularly polarized light. A natural extension would be to quasi-3D metasurfaces for 2D acoustic holography. While this study provides initial steps towards algorithmically optimized acoustoplasmonic metasurfaces, we anticipate, with excitement, future studies involving more complex acoustic responses as well as experimental validations of such designs. 

\begin{acknowledgments}
We wish to acknowledge the support of Zaid Haddadin for his assistance with MATLAB coding.
\end{acknowledgments}

\section*{Funding Sources}

J.E.H. acknowledges support from the Air Force Office of Scientific Research Young Investigator Research Program Grant (FA9550-23-1-0263). N. B. acknowledges support from the US Army Research Office (Grant No. W911NF-20-2-0182). L.V.P. acknowledges support from the David and Lucile Packard Foundation (2023-76165).

\section*{Conflict of Interest}

The authors have no conflicts to disclose.

\section*{Data Availability Statement}

The data that support the findings of this study are available from the corresponding author upon reasonable request. 

\clearpage

\appendix

\section{\label{sec:analytical}Analytical pressure derivation}

For the case of the lens, we analytically simulate nanospheres excited by a Gaussian optical pulse,
\begin{equation}
    I(t)=I_0e^{-\frac{(t-t_{peak})^2}{2\sigma^2}},
\end{equation}
where $I_0$ is the peak intensity magnitude, $t_{peak}$ is the time at peak intensity, and the standard deviation $\sigma = \tau_P/\left(2\sqrt{2ln2} \right) \approx \tau_P/2.355$.

For the case of a 20 ns pulse, the temperature rise in the nanosphere (assuming uniform temperature gradient) can be approximated as
\begin{equation}
    T(t)=\frac{C_{abs}I(t)}{4\pi R \kappa_f},
\end{equation}
where $R$ is the radius and $\kappa_f$ is the thermal conductivity coefficient of water.\cite{tribelsky2011laser} We arbitrarily choose an absorption efficiency of 1.4 and peak optical intensity of $5$ GW/m$^2$. 

The thermal expansion in the radial direction is modeled through the strain relationship
\begin{equation}
    \epsilon=\frac{u}{R}=\alpha(T-T_0),
\end{equation}
where $u$ is the displacement at the particle surface, $\alpha$ is the linear thermal expansion coefficient, and $T_0$ is the initial temperature.\cite{hibbeler2005mechanics}

The pressure is calculated from the radial surface displacement from
\begin{equation}
    \rho_f\frac{\partial^2u}{\partial t^2}=-\nabla p,
\end{equation}
where $\rho_f$ is the fluid density and $p$ is the pressure from equilibrium, where positive pressure is taken as compression.\cite{royer1999elastic} The relative phase shift between the displacement and pressure at the surface for diverging spherical waves is dictated by the interplay between the wavenumber $k$ and $R^{-1}$, which results in a $-\pi$ phase shift for the case of a long temporal pulse and small particle.\cite{kinsler2000fundamentals,royer1999elastic} For this very specific case, we derive an approximation for the pressure generation at the nanoparticle surface
\begin{equation}
    p=\rho_fR\frac{\partial^2u}{\partial t^2},
\end{equation}
which is shown in Figure \ref{fig:P0} (orange dashed line).

\section{\label{sec:numerical}Numerical simulation methods}

We numerically simulate gold nanoellipsoids using COMSOL Multiphysics 6.2, beginning with the wavelength-dependent optical absorption using the Electromagnetic Waves, Frequency Domain physics module.\cite{yushanov2013mie,COMSOL:MieScat} We use a 3D geometry space in which we model the particle in a spherical domain of fluid having a radius of 250 nm, and a perfectly matched layer of 250 nm. The mesh is left at physics-controlled and set to resolve the wave in lossy media (resolving the wave down to the skin depth of the gold particle). Peak absorption is found to occur at 916 nm (to 2 nm resolution).

In a separate simulation, we couple the Heat Transfer in Solids, Solid Mechanics, and Pressure Acoustics transient physics modules using Thermal Expansion and Acoustic-Structure Boundary multiphysics modules. We use quarter-symmetry in a 2D axisymmetric setup. The particle is positioned in a fluid layer of 1 $\mu$m thickness with a perfectly matched layer of 200 nm thickness. The maximum frequency to resolve ($f_{max}$) is set to $2/\tau_P$ and the solver time step is $dt = 1/\left(60 f_{max} \right)$. The maximum mesh element size is set to 5 nm. The particle is excited with a uniform heat generation rate density
\begin{equation}
    \dot{Q}(t)=\frac{C_{abs}I(t)}{V},
\end{equation}
where $V$ is the the volume.\cite{wang2022understanding} The pressure response is probed at the particle surface at the minor axis.

Similar simulations are described more comprehensively in the Supplementary Information in Ref\cite{holland2024acoustoplasmonic}.

\bibliography{aipsamp}

\providecommand{\noopsort}[1]{}\providecommand{\singleletter}[1]{#1}%
\begin{thebibliography}{30}%
\makeatletter
\providecommand \@ifxundefined [1]{%
 \@ifx{#1\undefined}
}%
\providecommand \@ifnum [1]{%
 \ifnum #1\expandafter \@firstoftwo
 \else \expandafter \@secondoftwo
 \fi
}%
\providecommand \@ifx [1]{%
 \ifx #1\expandafter \@firstoftwo
 \else \expandafter \@secondoftwo
 \fi
}%
\providecommand \natexlab [1]{#1}%
\providecommand \enquote  [1]{``#1''}%
\providecommand \bibnamefont  [1]{#1}%
\providecommand \bibfnamefont [1]{#1}%
\providecommand \citenamefont [1]{#1}%
\providecommand \href@noop [0]{\@secondoftwo}%
\providecommand \href [0]{\begingroup \@sanitize@url \@href}%
\providecommand \@href[1]{\@@startlink{#1}\@@href}%
\providecommand \@@href[1]{\endgroup#1\@@endlink}%
\providecommand \@sanitize@url [0]{\catcode `\\12\catcode `\$12\catcode `\&12\catcode `\#12\catcode `\^12\catcode `\_12\catcode `\%12\relax}%
\providecommand \@@startlink[1]{}%
\providecommand \@@endlink[0]{}%
\providecommand \url  [0]{\begingroup\@sanitize@url \@url }%
\providecommand \@url [1]{\endgroup\@href {#1}{\urlprefix }}%
\providecommand \urlprefix  [0]{URL }%
\providecommand \Eprint [0]{\href }%
\providecommand \doibase [0]{http://dx.doi.org/}%
\providecommand \selectlanguage [0]{\@gobble}%
\providecommand \bibinfo  [0]{\@secondoftwo}%
\providecommand \bibfield  [0]{\@secondoftwo}%
\providecommand \translation [1]{[#1]}%
\providecommand \BibitemOpen [0]{}%
\providecommand \bibitemStop [0]{}%
\providecommand \bibitemNoStop [0]{.\EOS\space}%
\providecommand \EOS [0]{\spacefactor3000\relax}%
\providecommand \BibitemShut  [1]{\csname bibitem#1\endcsname}%
\let\auto@bib@innerbib\@empty
\bibitem [{\citenamefont {Kuznetsov}\ \emph {et~al.}(2024)\citenamefont {Kuznetsov}, \citenamefont {Brongersma}, \citenamefont {Yao}, \citenamefont {Chen}, \citenamefont {Levy}, \citenamefont {Tsai}, \citenamefont {Zheludev}, \citenamefont {Faraon}, \citenamefont {Arbabi}, \citenamefont {Yu} \emph {et~al.}}]{kuznetsov2024roadmap}%
  \BibitemOpen
  \bibfield  {author} {\bibinfo {author} {\bibfnamefont {A.~I.}\ \bibnamefont {Kuznetsov}}, \bibinfo {author} {\bibfnamefont {M.~L.}\ \bibnamefont {Brongersma}}, \bibinfo {author} {\bibfnamefont {J.}~\bibnamefont {Yao}}, \bibinfo {author} {\bibfnamefont {M.~K.}\ \bibnamefont {Chen}}, \bibinfo {author} {\bibfnamefont {U.}~\bibnamefont {Levy}}, \bibinfo {author} {\bibfnamefont {D.~P.}\ \bibnamefont {Tsai}}, \bibinfo {author} {\bibfnamefont {N.~I.}\ \bibnamefont {Zheludev}}, \bibinfo {author} {\bibfnamefont {A.}~\bibnamefont {Faraon}}, \bibinfo {author} {\bibfnamefont {A.}~\bibnamefont {Arbabi}}, \bibinfo {author} {\bibfnamefont {N.}~\bibnamefont {Yu}},  \emph {et~al.},\ }\bibfield  {title} {\enquote {\bibinfo {title} {Roadmap for optical metasurfaces},}\ }\href@noop {} {\bibfield  {journal} {\bibinfo  {journal} {ACS photonics}\ }\textbf {\bibinfo {volume} {11}},\ \bibinfo {pages} {816--865} (\bibinfo {year} {2024})}\BibitemShut {NoStop}%
\bibitem [{\citenamefont {Assouar}\ \emph {et~al.}(2018)\citenamefont {Assouar}, \citenamefont {Liang}, \citenamefont {Wu}, \citenamefont {Li}, \citenamefont {Cheng},\ and\ \citenamefont {Jing}}]{assouar2018acoustic}%
  \BibitemOpen
  \bibfield  {author} {\bibinfo {author} {\bibfnamefont {B.}~\bibnamefont {Assouar}}, \bibinfo {author} {\bibfnamefont {B.}~\bibnamefont {Liang}}, \bibinfo {author} {\bibfnamefont {Y.}~\bibnamefont {Wu}}, \bibinfo {author} {\bibfnamefont {Y.}~\bibnamefont {Li}}, \bibinfo {author} {\bibfnamefont {J.-C.}\ \bibnamefont {Cheng}}, \ and\ \bibinfo {author} {\bibfnamefont {Y.}~\bibnamefont {Jing}},\ }\bibfield  {title} {\enquote {\bibinfo {title} {Acoustic metasurfaces},}\ }\href@noop {} {\bibfield  {journal} {\bibinfo  {journal} {Nature Reviews Materials}\ }\textbf {\bibinfo {volume} {3}},\ \bibinfo {pages} {460--472} (\bibinfo {year} {2018})}\BibitemShut {NoStop}%
\bibitem [{\citenamefont {Ji}\ and\ \citenamefont {Huber}(2022)}]{ji2022recent}%
  \BibitemOpen
  \bibfield  {author} {\bibinfo {author} {\bibfnamefont {G.}~\bibnamefont {Ji}}\ and\ \bibinfo {author} {\bibfnamefont {J.}~\bibnamefont {Huber}},\ }\bibfield  {title} {\enquote {\bibinfo {title} {Recent progress in acoustic metamaterials and active piezoelectric acoustic metamaterials-a review},}\ }\href@noop {} {\bibfield  {journal} {\bibinfo  {journal} {Applied Materials Today}\ }\textbf {\bibinfo {volume} {26}},\ \bibinfo {pages} {101260} (\bibinfo {year} {2022})}\BibitemShut {NoStop}%
\bibitem [{\citenamefont {Zou}\ \emph {et~al.}(2023)\citenamefont {Zou}, \citenamefont {Jin}, \citenamefont {Zhu},\ and\ \citenamefont {Zhang}}]{zou2023metasurface}%
  \BibitemOpen
  \bibfield  {author} {\bibinfo {author} {\bibfnamefont {Y.}~\bibnamefont {Zou}}, \bibinfo {author} {\bibfnamefont {H.}~\bibnamefont {Jin}}, \bibinfo {author} {\bibfnamefont {R.}~\bibnamefont {Zhu}}, \ and\ \bibinfo {author} {\bibfnamefont {T.}~\bibnamefont {Zhang}},\ }\bibfield  {title} {\enquote {\bibinfo {title} {Metasurface holography with multiplexing and reconfigurability},}\ }\href@noop {} {\bibfield  {journal} {\bibinfo  {journal} {Nanomaterials}\ }\textbf {\bibinfo {volume} {14}},\ \bibinfo {pages} {66} (\bibinfo {year} {2023})}\BibitemShut {NoStop}%
\bibitem [{\citenamefont {Elsawy}\ \emph {et~al.}(2020)\citenamefont {Elsawy}, \citenamefont {Lanteri}, \citenamefont {Duvigneau}, \citenamefont {Fan},\ and\ \citenamefont {Genevet}}]{elsawy2020numerical}%
  \BibitemOpen
  \bibfield  {author} {\bibinfo {author} {\bibfnamefont {M.~M.}\ \bibnamefont {Elsawy}}, \bibinfo {author} {\bibfnamefont {S.}~\bibnamefont {Lanteri}}, \bibinfo {author} {\bibfnamefont {R.}~\bibnamefont {Duvigneau}}, \bibinfo {author} {\bibfnamefont {J.~A.}\ \bibnamefont {Fan}}, \ and\ \bibinfo {author} {\bibfnamefont {P.}~\bibnamefont {Genevet}},\ }\bibfield  {title} {\enquote {\bibinfo {title} {Numerical optimization methods for metasurfaces},}\ }\href@noop {} {\bibfield  {journal} {\bibinfo  {journal} {Laser \& Photonics Reviews}\ }\textbf {\bibinfo {volume} {14}},\ \bibinfo {pages} {1900445} (\bibinfo {year} {2020})}\BibitemShut {NoStop}%
\bibitem [{\citenamefont {Melde}\ \emph {et~al.}(2016)\citenamefont {Melde}, \citenamefont {Mark}, \citenamefont {Qiu},\ and\ \citenamefont {Fischer}}]{melde2016holograms}%
  \BibitemOpen
  \bibfield  {author} {\bibinfo {author} {\bibfnamefont {K.}~\bibnamefont {Melde}}, \bibinfo {author} {\bibfnamefont {A.~G.}\ \bibnamefont {Mark}}, \bibinfo {author} {\bibfnamefont {T.}~\bibnamefont {Qiu}}, \ and\ \bibinfo {author} {\bibfnamefont {P.}~\bibnamefont {Fischer}},\ }\bibfield  {title} {\enquote {\bibinfo {title} {Holograms for acoustics},}\ }\href@noop {} {\bibfield  {journal} {\bibinfo  {journal} {Nature}\ }\textbf {\bibinfo {volume} {537}},\ \bibinfo {pages} {518--522} (\bibinfo {year} {2016})}\BibitemShut {NoStop}%
\bibitem [{\citenamefont {Miao}\ \emph {et~al.}(2023)\citenamefont {Miao}, \citenamefont {Dong}, \citenamefont {Zhao}, \citenamefont {Fan}, \citenamefont {Huang}, \citenamefont {Shen},\ and\ \citenamefont {Wang}}]{miao2023deep}%
  \BibitemOpen
  \bibfield  {author} {\bibinfo {author} {\bibfnamefont {X.-B.}\ \bibnamefont {Miao}}, \bibinfo {author} {\bibfnamefont {H.-W.}\ \bibnamefont {Dong}}, \bibinfo {author} {\bibfnamefont {S.-D.}\ \bibnamefont {Zhao}}, \bibinfo {author} {\bibfnamefont {S.-W.}\ \bibnamefont {Fan}}, \bibinfo {author} {\bibfnamefont {G.}~\bibnamefont {Huang}}, \bibinfo {author} {\bibfnamefont {C.}~\bibnamefont {Shen}}, \ and\ \bibinfo {author} {\bibfnamefont {Y.-S.}\ \bibnamefont {Wang}},\ }\bibfield  {title} {\enquote {\bibinfo {title} {Deep-learning-aided metasurface design for megapixel acoustic hologram},}\ }\href@noop {} {\bibfield  {journal} {\bibinfo  {journal} {Applied Physics Reviews}\ }\textbf {\bibinfo {volume} {10}} (\bibinfo {year} {2023})}\BibitemShut {NoStop}%
\bibitem [{\citenamefont {Liu}\ and\ \citenamefont {Waag}(1997)}]{liu1997propagation}%
  \BibitemOpen
  \bibfield  {author} {\bibinfo {author} {\bibfnamefont {D.-L.}\ \bibnamefont {Liu}}\ and\ \bibinfo {author} {\bibfnamefont {R.~C.}\ \bibnamefont {Waag}},\ }\bibfield  {title} {\enquote {\bibinfo {title} {Propagation and backpropagation for ultrasonic wavefront design},}\ }\href@noop {} {\bibfield  {journal} {\bibinfo  {journal} {IEEE transactions on ultrasonics, ferroelectrics, and frequency control}\ }\textbf {\bibinfo {volume} {44}},\ \bibinfo {pages} {1--13} (\bibinfo {year} {1997})}\BibitemShut {NoStop}%
\bibitem [{\citenamefont {Fayyaz}\ \emph {et~al.}(2019)\citenamefont {Fayyaz}, \citenamefont {Mohammadian}, \citenamefont {Reza Rahimi~Tabar}, \citenamefont {Manwar},\ and\ \citenamefont {Avanaki}}]{fayyaz2019comparative}%
  \BibitemOpen
  \bibfield  {author} {\bibinfo {author} {\bibfnamefont {Z.}~\bibnamefont {Fayyaz}}, \bibinfo {author} {\bibfnamefont {N.}~\bibnamefont {Mohammadian}}, \bibinfo {author} {\bibfnamefont {M.}~\bibnamefont {Reza Rahimi~Tabar}}, \bibinfo {author} {\bibfnamefont {R.}~\bibnamefont {Manwar}}, \ and\ \bibinfo {author} {\bibfnamefont {K.}~\bibnamefont {Avanaki}},\ }\bibfield  {title} {\enquote {\bibinfo {title} {A comparative study of optimization algorithms for wavefront shaping},}\ }\href@noop {} {\bibfield  {journal} {\bibinfo  {journal} {Journal of innovative optical health sciences}\ }\textbf {\bibinfo {volume} {12}},\ \bibinfo {pages} {1942002} (\bibinfo {year} {2019})}\BibitemShut {NoStop}%
\bibitem [{\citenamefont {Haim}, \citenamefont {Boger-Lombard},\ and\ \citenamefont {Katz}(2024)}]{haim2024image}%
  \BibitemOpen
  \bibfield  {author} {\bibinfo {author} {\bibfnamefont {O.}~\bibnamefont {Haim}}, \bibinfo {author} {\bibfnamefont {J.}~\bibnamefont {Boger-Lombard}}, \ and\ \bibinfo {author} {\bibfnamefont {O.}~\bibnamefont {Katz}},\ }\bibfield  {title} {\enquote {\bibinfo {title} {Image-guided computational holographic wavefront shaping},}\ }\href@noop {} {\bibfield  {journal} {\bibinfo  {journal} {Nature Photonics}\ ,\ \bibinfo {pages} {1--10}} (\bibinfo {year} {2024})}\BibitemShut {NoStop}%
\bibitem [{\citenamefont {Haddadin}\ and\ \citenamefont {Ebbini}(1998)}]{haddadin1998ultrasonic}%
  \BibitemOpen
  \bibfield  {author} {\bibinfo {author} {\bibfnamefont {O.~S.}\ \bibnamefont {Haddadin}}\ and\ \bibinfo {author} {\bibfnamefont {E.~S.}\ \bibnamefont {Ebbini}},\ }\bibfield  {title} {\enquote {\bibinfo {title} {Ultrasonic focusing through inhomogeneous media by application of the inverse scattering problem},}\ }\href@noop {} {\bibfield  {journal} {\bibinfo  {journal} {The Journal of the Acoustical Society of America}\ }\textbf {\bibinfo {volume} {104}},\ \bibinfo {pages} {313--325} (\bibinfo {year} {1998})}\BibitemShut {NoStop}%
\bibitem [{\citenamefont {Berte}\ \emph {et~al.}(2018)\citenamefont {Berte}, \citenamefont {Della~Picca}, \citenamefont {Poblet}, \citenamefont {Li}, \citenamefont {Cort{\'e}s}, \citenamefont {Craster}, \citenamefont {Maier},\ and\ \citenamefont {Bragas}}]{berte2018acoustic}%
  \BibitemOpen
  \bibfield  {author} {\bibinfo {author} {\bibfnamefont {R.}~\bibnamefont {Berte}}, \bibinfo {author} {\bibfnamefont {F.}~\bibnamefont {Della~Picca}}, \bibinfo {author} {\bibfnamefont {M.}~\bibnamefont {Poblet}}, \bibinfo {author} {\bibfnamefont {Y.}~\bibnamefont {Li}}, \bibinfo {author} {\bibfnamefont {E.}~\bibnamefont {Cort{\'e}s}}, \bibinfo {author} {\bibfnamefont {R.~V.}\ \bibnamefont {Craster}}, \bibinfo {author} {\bibfnamefont {S.~A.}\ \bibnamefont {Maier}}, \ and\ \bibinfo {author} {\bibfnamefont {A.~V.}\ \bibnamefont {Bragas}},\ }\bibfield  {title} {\enquote {\bibinfo {title} {Acoustic far-field hypersonic surface wave detection with single plasmonic nanoantennas},}\ }\href@noop {} {\bibfield  {journal} {\bibinfo  {journal} {Physical Review Letters}\ }\textbf {\bibinfo {volume} {121}},\ \bibinfo {pages} {253902} (\bibinfo {year} {2018})}\BibitemShut {NoStop}%
\bibitem [{\citenamefont {Imade}\ \emph {et~al.}(2021)\citenamefont {Imade}, \citenamefont {Gusev}, \citenamefont {Matsuda}, \citenamefont {Tomoda}, \citenamefont {Otsuka},\ and\ \citenamefont {Wright}}]{imade2021gigahertz}%
  \BibitemOpen
  \bibfield  {author} {\bibinfo {author} {\bibfnamefont {Y.}~\bibnamefont {Imade}}, \bibinfo {author} {\bibfnamefont {V.~E.}\ \bibnamefont {Gusev}}, \bibinfo {author} {\bibfnamefont {O.}~\bibnamefont {Matsuda}}, \bibinfo {author} {\bibfnamefont {M.}~\bibnamefont {Tomoda}}, \bibinfo {author} {\bibfnamefont {P.~H.}\ \bibnamefont {Otsuka}}, \ and\ \bibinfo {author} {\bibfnamefont {O.~B.}\ \bibnamefont {Wright}},\ }\bibfield  {title} {\enquote {\bibinfo {title} {Gigahertz optomechanical photon--phonon transduction between nanostructure lines},}\ }\href@noop {} {\bibfield  {journal} {\bibinfo  {journal} {Nano Letters}\ }\textbf {\bibinfo {volume} {21}},\ \bibinfo {pages} {6261--6267} (\bibinfo {year} {2021})}\BibitemShut {NoStop}%
\bibitem [{\citenamefont {Poblet}\ \emph {et~al.}(2021)\citenamefont {Poblet}, \citenamefont {Bert{\'e}}, \citenamefont {Boggiano}, \citenamefont {Li}, \citenamefont {Cort{\'e}s}, \citenamefont {Grinblat}, \citenamefont {Maier},\ and\ \citenamefont {Bragas}}]{poblet2021acoustic}%
  \BibitemOpen
  \bibfield  {author} {\bibinfo {author} {\bibfnamefont {M.}~\bibnamefont {Poblet}}, \bibinfo {author} {\bibfnamefont {R.}~\bibnamefont {Bert{\'e}}}, \bibinfo {author} {\bibfnamefont {H.~D.}\ \bibnamefont {Boggiano}}, \bibinfo {author} {\bibfnamefont {Y.}~\bibnamefont {Li}}, \bibinfo {author} {\bibfnamefont {E.}~\bibnamefont {Cort{\'e}s}}, \bibinfo {author} {\bibfnamefont {G.}~\bibnamefont {Grinblat}}, \bibinfo {author} {\bibfnamefont {S.~A.}\ \bibnamefont {Maier}}, \ and\ \bibinfo {author} {\bibfnamefont {A.~V.}\ \bibnamefont {Bragas}},\ }\bibfield  {title} {\enquote {\bibinfo {title} {Acoustic coupling between plasmonic nanoantennas: detection and directionality of surface acoustic waves},}\ }\href@noop {} {\bibfield  {journal} {\bibinfo  {journal} {ACS Photonics}\ }\textbf {\bibinfo {volume} {8}},\ \bibinfo {pages} {2846--2852} (\bibinfo {year} {2021})}\BibitemShut {NoStop}%
\bibitem [{\citenamefont {Boggiano}\ \emph {et~al.}(2024)\citenamefont {Boggiano}, \citenamefont {Nan}, \citenamefont {Grinblat}, \citenamefont {Maier}, \citenamefont {Cort{\'e}s},\ and\ \citenamefont {Bragas}}]{boggiano2024focusing}%
  \BibitemOpen
  \bibfield  {author} {\bibinfo {author} {\bibfnamefont {H.~D.}\ \bibnamefont {Boggiano}}, \bibinfo {author} {\bibfnamefont {L.}~\bibnamefont {Nan}}, \bibinfo {author} {\bibfnamefont {G.}~\bibnamefont {Grinblat}}, \bibinfo {author} {\bibfnamefont {S.~A.}\ \bibnamefont {Maier}}, \bibinfo {author} {\bibfnamefont {E.}~\bibnamefont {Cort{\'e}s}}, \ and\ \bibinfo {author} {\bibfnamefont {A.~V.}\ \bibnamefont {Bragas}},\ }\bibfield  {title} {\enquote {\bibinfo {title} {Focusing surface acoustic waves with a plasmonic hypersonic lens},}\ }\href@noop {} {\bibfield  {journal} {\bibinfo  {journal} {Nano Letters}\ } (\bibinfo {year} {2024})}\BibitemShut {NoStop}%
\bibitem [{\citenamefont {Drain}(2019)}]{drain2019laser}%
  \BibitemOpen
  \bibfield  {author} {\bibinfo {author} {\bibfnamefont {L.}~\bibnamefont {Drain}},\ }\href@noop {} {\emph {\bibinfo {title} {Laser ultrasonics techniques and applications}}}\ (\bibinfo  {publisher} {Routledge},\ \bibinfo {year} {2019})\BibitemShut {NoStop}%
\bibitem [{\citenamefont {Temnov}\ \emph {et~al.}(2016)\citenamefont {Temnov}, \citenamefont {Razdolski}, \citenamefont {Pezeril}, \citenamefont {Makarov}, \citenamefont {Seletskiy}, \citenamefont {Melnikov},\ and\ \citenamefont {Nelson}}]{temnov2016towards}%
  \BibitemOpen
  \bibfield  {author} {\bibinfo {author} {\bibfnamefont {V.~V.}\ \bibnamefont {Temnov}}, \bibinfo {author} {\bibfnamefont {I.}~\bibnamefont {Razdolski}}, \bibinfo {author} {\bibfnamefont {T.}~\bibnamefont {Pezeril}}, \bibinfo {author} {\bibfnamefont {D.}~\bibnamefont {Makarov}}, \bibinfo {author} {\bibfnamefont {D.}~\bibnamefont {Seletskiy}}, \bibinfo {author} {\bibfnamefont {A.}~\bibnamefont {Melnikov}}, \ and\ \bibinfo {author} {\bibfnamefont {K.~A.}\ \bibnamefont {Nelson}},\ }\bibfield  {title} {\enquote {\bibinfo {title} {Towards the nonlinear acousto-magneto-plasmonics},}\ }\href@noop {} {\bibfield  {journal} {\bibinfo  {journal} {Journal of Optics}\ }\textbf {\bibinfo {volume} {18}},\ \bibinfo {pages} {093002} (\bibinfo {year} {2016})}\BibitemShut {NoStop}%
\bibitem [{\citenamefont {Cielo}, \citenamefont {Nadeau},\ and\ \citenamefont {Lamontagne}(1985)}]{cielo1985laser}%
  \BibitemOpen
  \bibfield  {author} {\bibinfo {author} {\bibfnamefont {P.}~\bibnamefont {Cielo}}, \bibinfo {author} {\bibfnamefont {F.}~\bibnamefont {Nadeau}}, \ and\ \bibinfo {author} {\bibfnamefont {M.}~\bibnamefont {Lamontagne}},\ }\bibfield  {title} {\enquote {\bibinfo {title} {Laser generation of convergent acoustic waves for materials inspection},}\ }\href@noop {} {\bibfield  {journal} {\bibinfo  {journal} {Ultrasonics}\ }\textbf {\bibinfo {volume} {23}},\ \bibinfo {pages} {55--62} (\bibinfo {year} {1985})}\BibitemShut {NoStop}%
\bibitem [{\citenamefont {Bruno}\ \emph {et~al.}(2016)\citenamefont {Bruno}, \citenamefont {Laurent}, \citenamefont {Jehanno}, \citenamefont {Royer},\ and\ \citenamefont {Prada}}]{bruno2016laser}%
  \BibitemOpen
  \bibfield  {author} {\bibinfo {author} {\bibfnamefont {F.}~\bibnamefont {Bruno}}, \bibinfo {author} {\bibfnamefont {J.}~\bibnamefont {Laurent}}, \bibinfo {author} {\bibfnamefont {P.}~\bibnamefont {Jehanno}}, \bibinfo {author} {\bibfnamefont {D.}~\bibnamefont {Royer}}, \ and\ \bibinfo {author} {\bibfnamefont {C.}~\bibnamefont {Prada}},\ }\bibfield  {title} {\enquote {\bibinfo {title} {Laser beam shaping for enhanced zero-group velocity lamb modes generation},}\ }\href@noop {} {\bibfield  {journal} {\bibinfo  {journal} {The Journal of the Acoustical Society of America}\ }\textbf {\bibinfo {volume} {140}},\ \bibinfo {pages} {2829--2838} (\bibinfo {year} {2016})}\BibitemShut {NoStop}%
\bibitem [{\citenamefont {Matsuda}\ \emph {et~al.}(2020)\citenamefont {Matsuda}, \citenamefont {Tsutsui}, \citenamefont {Vaudel}, \citenamefont {Pezeril}, \citenamefont {Fujita},\ and\ \citenamefont {Gusev}}]{matsuda2020optical}%
  \BibitemOpen
  \bibfield  {author} {\bibinfo {author} {\bibfnamefont {O.}~\bibnamefont {Matsuda}}, \bibinfo {author} {\bibfnamefont {K.}~\bibnamefont {Tsutsui}}, \bibinfo {author} {\bibfnamefont {G.}~\bibnamefont {Vaudel}}, \bibinfo {author} {\bibfnamefont {T.}~\bibnamefont {Pezeril}}, \bibinfo {author} {\bibfnamefont {K.}~\bibnamefont {Fujita}}, \ and\ \bibinfo {author} {\bibfnamefont {V.}~\bibnamefont {Gusev}},\ }\bibfield  {title} {\enquote {\bibinfo {title} {Optical generation and detection of gigahertz shear acoustic waves in solids assisted by a metallic diffraction grating},}\ }\href@noop {} {\bibfield  {journal} {\bibinfo  {journal} {Physical Review B}\ }\textbf {\bibinfo {volume} {101}},\ \bibinfo {pages} {224307} (\bibinfo {year} {2020})}\BibitemShut {NoStop}%
\bibitem [{\citenamefont {Holland}\ \emph {et~al.}(2024)\citenamefont {Holland}, \citenamefont {Byun}, \citenamefont {Boechler},\ and\ \citenamefont {Poulikakos}}]{holland2024acoustoplasmonic}%
  \BibitemOpen
  \bibfield  {author} {\bibinfo {author} {\bibfnamefont {J.~E.}\ \bibnamefont {Holland}}, \bibinfo {author} {\bibfnamefont {J.}~\bibnamefont {Byun}}, \bibinfo {author} {\bibfnamefont {N.}~\bibnamefont {Boechler}}, \ and\ \bibinfo {author} {\bibfnamefont {L.~V.}\ \bibnamefont {Poulikakos}},\ }\bibfield  {title} {\enquote {\bibinfo {title} {Acoustoplasmonic metasurfaces for tunable acoustic wavefront shaping with polarized light},}\ }\href@noop {} {\bibfield  {journal} {\bibinfo  {journal} {ACS Photonics}\ } (\bibinfo {year} {2024})}\BibitemShut {NoStop}%
\bibitem [{\citenamefont {Kinsler}\ \emph {et~al.}(2000)\citenamefont {Kinsler}, \citenamefont {Frey}, \citenamefont {Coppens},\ and\ \citenamefont {Sanders}}]{kinsler2000fundamentals}%
  \BibitemOpen
  \bibfield  {author} {\bibinfo {author} {\bibfnamefont {L.~E.}\ \bibnamefont {Kinsler}}, \bibinfo {author} {\bibfnamefont {A.~R.}\ \bibnamefont {Frey}}, \bibinfo {author} {\bibfnamefont {A.~B.}\ \bibnamefont {Coppens}}, \ and\ \bibinfo {author} {\bibfnamefont {J.~V.}\ \bibnamefont {Sanders}},\ }\href@noop {} {\emph {\bibinfo {title} {Fundamentals of acoustics}}}\ (\bibinfo  {publisher} {John wiley \& sons},\ \bibinfo {year} {2000})\BibitemShut {NoStop}%
\bibitem [{\citenamefont {Hecht}(2012)}]{hecht2012optics}%
  \BibitemOpen
  \bibfield  {author} {\bibinfo {author} {\bibfnamefont {E.}~\bibnamefont {Hecht}},\ }\href@noop {} {\emph {\bibinfo {title} {Optics}}}\ (\bibinfo  {publisher} {Pearson Education India},\ \bibinfo {year} {2012})\BibitemShut {NoStop}%
\bibitem [{\citenamefont {Lanzillotti-Kimura}\ \emph {et~al.}(2018)\citenamefont {Lanzillotti-Kimura}, \citenamefont {O’Brien}, \citenamefont {Rho}, \citenamefont {Suchowski}, \citenamefont {Yin},\ and\ \citenamefont {Zhang}}]{lanzillotti2018polarization}%
  \BibitemOpen
  \bibfield  {author} {\bibinfo {author} {\bibfnamefont {N.~D.}\ \bibnamefont {Lanzillotti-Kimura}}, \bibinfo {author} {\bibfnamefont {K.~P.}\ \bibnamefont {O’Brien}}, \bibinfo {author} {\bibfnamefont {J.}~\bibnamefont {Rho}}, \bibinfo {author} {\bibfnamefont {H.}~\bibnamefont {Suchowski}}, \bibinfo {author} {\bibfnamefont {X.}~\bibnamefont {Yin}}, \ and\ \bibinfo {author} {\bibfnamefont {X.}~\bibnamefont {Zhang}},\ }\bibfield  {title} {\enquote {\bibinfo {title} {Polarization-controlled coherent phonon generation in acoustoplasmonic metasurfaces},}\ }\href@noop {} {\bibfield  {journal} {\bibinfo  {journal} {Physical Review B}\ }\textbf {\bibinfo {volume} {97}},\ \bibinfo {pages} {235403} (\bibinfo {year} {2018})}\BibitemShut {NoStop}%
\bibitem [{\citenamefont {Wang}, \citenamefont {Chen},\ and\ \citenamefont {Zhao}(2022)}]{wang2022understanding}%
  \BibitemOpen
  \bibfield  {author} {\bibinfo {author} {\bibfnamefont {H.}~\bibnamefont {Wang}}, \bibinfo {author} {\bibfnamefont {Y.-S.}\ \bibnamefont {Chen}}, \ and\ \bibinfo {author} {\bibfnamefont {Y.}~\bibnamefont {Zhao}},\ }\bibfield  {title} {\enquote {\bibinfo {title} {Understanding the near-field photoacoustic spatiotemporal profile from nanostructures},}\ }\href@noop {} {\bibfield  {journal} {\bibinfo  {journal} {Photoacoustics}\ }\textbf {\bibinfo {volume} {28}},\ \bibinfo {pages} {100425} (\bibinfo {year} {2022})}\BibitemShut {NoStop}%
\bibitem [{\citenamefont {Tribelsky}\ \emph {et~al.}(2011)\citenamefont {Tribelsky}, \citenamefont {Miroshnichenko}, \citenamefont {Kivshar}, \citenamefont {Luk’yanchuk},\ and\ \citenamefont {Khokhlov}}]{tribelsky2011laser}%
  \BibitemOpen
  \bibfield  {author} {\bibinfo {author} {\bibfnamefont {M.~I.}\ \bibnamefont {Tribelsky}}, \bibinfo {author} {\bibfnamefont {A.~E.}\ \bibnamefont {Miroshnichenko}}, \bibinfo {author} {\bibfnamefont {Y.~S.}\ \bibnamefont {Kivshar}}, \bibinfo {author} {\bibfnamefont {B.~S.}\ \bibnamefont {Luk’yanchuk}}, \ and\ \bibinfo {author} {\bibfnamefont {A.~R.}\ \bibnamefont {Khokhlov}},\ }\bibfield  {title} {\enquote {\bibinfo {title} {Laser pulse heating of spherical metal particles},}\ }\href@noop {} {\bibfield  {journal} {\bibinfo  {journal} {Physical Review X}\ }\textbf {\bibinfo {volume} {1}},\ \bibinfo {pages} {021024} (\bibinfo {year} {2011})}\BibitemShut {NoStop}%
\bibitem [{\citenamefont {Hibbeler}(2005)}]{hibbeler2005mechanics}%
  \BibitemOpen
  \bibfield  {author} {\bibinfo {author} {\bibfnamefont {R.~C.}\ \bibnamefont {Hibbeler}},\ }\href@noop {} {\emph {\bibinfo {title} {Mechanics of materials}}}\ (\bibinfo  {publisher} {Pearson Education India},\ \bibinfo {year} {2005})\BibitemShut {NoStop}%
\bibitem [{\citenamefont {Royer}\ and\ \citenamefont {Dieulesaint}(1999)}]{royer1999elastic}%
  \BibitemOpen
  \bibfield  {author} {\bibinfo {author} {\bibfnamefont {D.}~\bibnamefont {Royer}}\ and\ \bibinfo {author} {\bibfnamefont {E.}~\bibnamefont {Dieulesaint}},\ }\href@noop {} {\emph {\bibinfo {title} {Elastic waves in solids I: Free and guided propagation}}}\ (\bibinfo  {publisher} {Springer Science \& Business Media},\ \bibinfo {year} {1999})\BibitemShut {NoStop}%
\bibitem [{\citenamefont {Yushanov}, \citenamefont {Crompton},\ and\ \citenamefont {Koppenhoefer}(2013)}]{yushanov2013mie}%
  \BibitemOpen
  \bibfield  {author} {\bibinfo {author} {\bibfnamefont {S.}~\bibnamefont {Yushanov}}, \bibinfo {author} {\bibfnamefont {J.~S.}\ \bibnamefont {Crompton}}, \ and\ \bibinfo {author} {\bibfnamefont {K.~C.}\ \bibnamefont {Koppenhoefer}},\ }\bibfield  {title} {\enquote {\bibinfo {title} {Mie scattering of electromagnetic waves},}\ }in\ \href@noop {} {\emph {\bibinfo {booktitle} {Proceedings of the COMSOL Conference}}},\ Vol.\ \bibinfo {volume} {116}\ (\bibinfo {organization} {Boston},\ \bibinfo {year} {2013})\ pp.\ \bibinfo {pages} {1--7}\BibitemShut {NoStop}%
\bibitem [{COM()}]{COMSOL:MieScat}%
  \BibitemOpen
  \href {https://doc.comsol.com/5.5/doc/com.comsol.help.models.woptics.scattering_nanosphere/models.woptics.scattering_nanosphere.pdf} {\enquote {\bibinfo {title} {Optical scattering off a gold nanosphere},}\ }\bibinfo {note} {(Accessed: 2025-01-13)}\BibitemShut {NoStop}%
\end{thebibliography}%

\end{document}